\documentclass[aps,prb, twocolumn, superscriptaddress]{revtex4}

\usepackage{graphicx}
\usepackage{dcolumn}
\usepackage{bm}
\usepackage{xspace, epstopdf}
\usepackage{amsmath}    
\newcommand{\traj}{Fig.~\ref{traj}}
\newcommand{\disorder}{Fig.~\ref{disorder}}

\newcommand{\mfp}{Fig.~\ref{mfp}}

\begin{document}


\title{Numerical study of resonant spin relaxation in quasi-1D channels}
\author{S. L\"uscher}
\affiliation{Department of Physics and Astronomy, University of British Columbia, Vancouver, BC V6T 1Z4, Canada}
\author{S. M. Frolov}
\affiliation{Department of Physics and Astronomy, University of British Columbia, Vancouver, BC V6T 1Z4, Canada}
\affiliation{Kavli Institute of Nanoscience, Delft University of Technology, 2600 GA Delft, The Netherlands}
\author{J. A. Folk}
\affiliation{Department of Physics and Astronomy, University of British Columbia, Vancouver, BC V6T 1Z4, Canada}
\date{\today}
\begin{abstract}

Recent experiments demonstrate that a ballistic version of spin resonance, mediated by spin-orbit interaction, can be induced in narrow channels of a high-mobility GaAs two-dimensional electron gas by matching the spin precession frequency with the frequency of bouncing trajectories in the channel.  Contrary to the typical suppression of  Dyakonov-Perel' spin relaxation in confined geometries, the spin relaxation rate increases by orders of magnitude on resonance. Here, we present Monte Carlo simulations of this effect to explore the roles of varying degrees of disorder and strength of spin-orbit interaction.  These simulations help to extract quantitative spin-orbit parameters from experimental measurements of ballistic spin resonance, and may guide the development of future spintronic devices.
\end{abstract}

\maketitle

 \section{Introduction}

Spin-orbit interaction is the primary source of spin relaxation for free carriers in semiconductors.\cite{Pikus:1984, ZuticRMP}   At the same time, it offers the potential to control the spin orientation of those carriers without the need for conventional high frequency resonance techniques.\cite{BardarsonPRL07,Datta:1990, frolov_nature} Controlling carrier spins using spin-orbit interaction requires the ability to tune its effect with external parameters such as a magnetic field or voltages on electrostatic gates. In the Datta-Das spin transistor concept, for example, the spins of carriers in a 2D quantum well rotate in response to a spin-orbit interaction whose strength can be tuned by a gate.\cite{Datta:1990, Koo:2009}

Another way that electrostatic gates can tune the effects of spin-orbit interaction in a quantum well is by defining the lateral confinement geometry of a spintronic device.   Recent experiments have shown that bouncing trajectories in gate-defined channels of high mobility GaAs two-dimensional electron gas (2DEG), in an external magnetic field, lead to rapid spin relaxation through a process we refer to as ballistic spin resonance (BSR).\cite{frolov_nature}  On resonance the effect of spin-orbit interaction is amplified by matching the bouncing frequency to the Larmor precession frequency, and the bouncing frequency depends on the gate-defined channel width. Although the mechanism of BSR is straightforward, it is not obvious that the effect should be visible for realistic parameters in a practical device.

In this report, semi-classical Monte Carlo simulations of spin dynamics are used to test the resilience of BSR over a wide range of device parameters.   The simulation models varying degrees and types of disorder, confinement potential from the electrostatic gates, and lack of perfect specularity on scattering off the channel walls.  We restrict our attention to electron-doped GaAs 2DEGs at low temperature, where the Dyakonov-Perel' mechanism has been shown to be the dominant source of spin relaxation.\cite{Pikus:1984, ZuticRMP}  A range of spin-orbit interaction strengths are explored in the simulation, including linear Rashba and Dresselhaus terms as well as the cubic Dresselhaus term.  BSR is found to be robust over a wide range of experimentally-accessible parameters, and not to depend sensitively on specific model of disorder.

\begin{figure*} 
 \includegraphics[width= 0.95\textwidth]{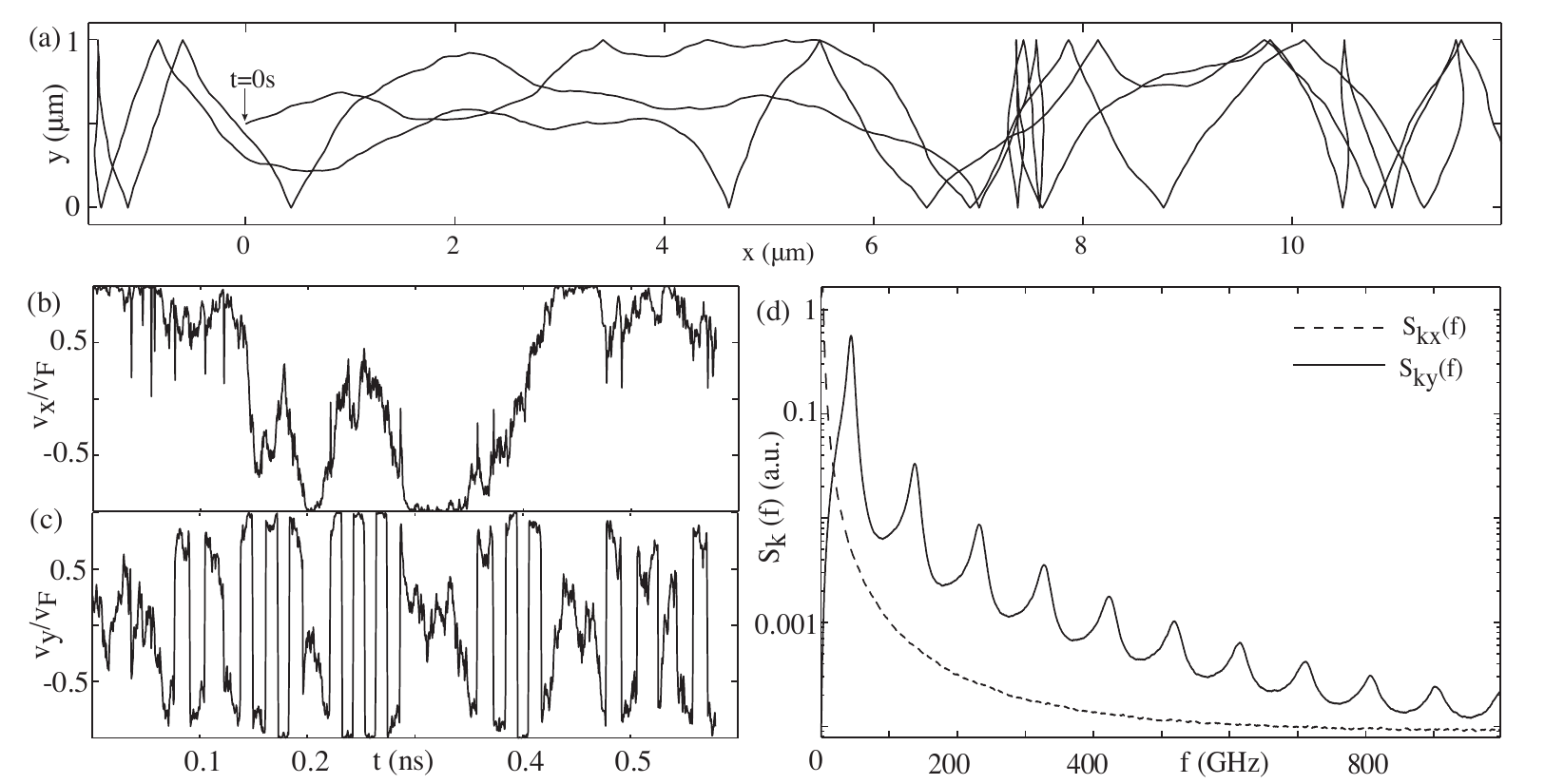}


\caption{\label{traj} {\bf (a)} An example of a trajectory segment in a $1\mu$m-wide wire with $\lambda$=10$\mu$m due to small angle scattering. Because the channel width is much less than the mean free path, the time dependence of the x- and y-components of the velocity, $v_x$ and $v_y$ respectively, are qualitatively different (panels {\bf (b)} and {\bf (c)}).  {\bf (d)}  Power spectral densities show the square wave nature of $k_y\propto v_y$ but no characteristic frequencies for $k_x \propto v_x$.}\end{figure*}

\section{Dyakonov-Perel' mechanism}

At the most general level, spin-orbit interaction couples an electron's spin degree of freedom to its momentum.
Spin-orbit interaction in III-V semiconductor quantum wells is characterized by Rashba ($\alpha$) and Dresselhaus ($\beta$ and $\gamma$) terms, due respectively to structural and bulk crystal inversion asymmetry.\cite{ZuticRMP}  Including both types of spin-orbit interaction, the spin-orbit Hamiltonian is:
\begin{eqnarray}
\label{eq:hso}
  H_{so} & = & \alpha(k_{010}\sigma_{100}-k_{100}\sigma_{010})+\beta(k_{100}\sigma_{100}-k_{010}\sigma_{010}) \nonumber \\  & & { } +  \gamma(k_{010}^2 k_{100}\sigma_{100}-k_{100}^2 k_{010}\sigma_{010})
\end{eqnarray}
where $k_{100}$ is the component of the Fermi wavevector along the [100] crystal axis, $\sigma_{100}$ is the Pauli spin operator along the [100] axis.  We note that the sign convention adopted for this Hamiltonian gives $\beta\approx-\gamma<k_{001}^2>$.  \cite{Winkler, Krich2007}

The linear-in-$k$ terms in the Hamiltonian become simpler when described along [110] and $[\overline{1}10]$ axes.  For the rest of this paper, the [110] axis is referred to as the $x$-axis;  the $[\overline{1}10]$ axis is referred to as the $y$-axis.  $H_{so}$ can be interpreted in terms of a momentum-dependent effective Zeeman field, $\vec{B}_{so}$, which takes the following form when expressed along $x$- and $y$-axes:
\begin{equation}
\label{eq:bso}
\vec{B}_{so}=\frac{2}{g\mu_B}((\alpha-\beta)k_y\hat{x}-(\alpha+\beta)k_x\hat{y}) + O(k^3),
\end{equation}
This effective field corresponds to a spin-orbit precession time $\tau_{so}=\hbar/g\mu_B B_{so}$, where $g=0.44$ is the Land\'e g-factor in GaAs.

Spin precession according to $H_{so}$ is coherent from a microscopic point of view.  At a practical level, however, this term gives rise to spin relaxation in any real conductor due to momentum scattering.   Electron spins precess around an effective magnetic field that changes, as the momentum changes, at each scattering event.  An ensemble of polarized spins, initially oriented in the same direction but following different random trajectories, will be distributed randomly around the Bloch sphere after a relaxation time, $\tau_{sr}$.  This relaxation process is known as the Dyakonov-Perel' (DP) mechanism.\cite{DyakPerel}

External magnetic fields have a strong effect on DP relaxation.  These effects can be quite complicated when both orbital and spin effects are included.  In this paper we discuss only the case of in-plane magnetic fields, which give rise to Zeeman splitting but not to Landau quantization or cyclotron motion.   When $\vec{B}_{so}$ is added to an in-plane magnetic field $\vec{B}_{ext}$, it is the total effective field $\vec{B}_{tot} =\vec{B}_{so}+ \vec{B}_{ext}$ that sets the spin precession axis and precession time for DP spin dynamics.  The relaxation time, $\tau_{sr}$, due to the DP mechanism has been calculated\cite{Kiselev:2000} for disordered 2D systems with momentum scattering time $\tau_p$, giving 
\begin{equation}
\label{dp2d}
\tau_{sr}(B_{ext})\sim\frac{\tau_{so}^2}{\tau_p}(1+(\tau_p\frac{g\mu_B B_{ext}}{\hbar})^2).
\end{equation}  Here $\tau_p$ corresponds to a mean free path $\lambda=\tau_p v_F$.

The monotonic dependence $\tau_{sr}(B_{ext})$ described by Eq.~(\ref{dp2d}) does not hold in confined geometries, such as the channels studied here, where the mean free path and spin-orbit length are on the order of or greater than the channel width. \cite{HolleitnerPRL06} It is the goal of this work to study $\tau_{sr}(B_{ext})$ numerically in these cases.

\section{Model}

Semi-classical Monte Carlo simulations of DP spin dynamics in a 2DEG channel were performed by calculating momentum-dependent spin precession along an ensemble of randomly-generated classical trajectories, $\vec{r}_i(t)$, analogous to the calculations described in Refs.~\onlinecite{Kiselev:2000, Koop, Liu}.  Instantaneous velocity was determined from $\vec{v}_i(t)=d\vec{r}_i/dt$, giving momentum $\hbar\vec{k}_i(t)=m^*\vec{v}_i(t)$ for effective mass $m^*$.  Throughout this paper the magnitude of the velocity was $|\vec{v}_i|=v_F=10^5$m/s, corresponding to electron sheet density $n_s\approx5\times 10^{10} cm^{-2}$.\cite{frolov_prl, frolov_nature}

Each spin $\vec{s}_i$ was initialized to lie along the external field, $\vec{s}_i(t=0)\parallel\vec{B}_{ext}$.  The spins evolved in time by precessing around the trajectory-dependent $\vec{B}_{tot}(t)$, calculated using Eq.~(\ref{eq:bso}) and $\vec{k}_i(t)$.    The ensemble-averaged projection of the spin on the initial axis was then calculated as a function of time, $P(t)=\langle\vec{s}_i(t)\cdot\vec{s}_i(0)\rangle_i$, and fit to an exponential decay model, $P(t)=P_0 e^{-t/\tau_{sr}}$, to extract the spin relaxation time $\tau_{sr}$.  Although the DP mechanism can give rise to an oscillatory behavior for $P(t)$ when the external magnetic field is zero (data not shown),  the oscillatory component in $P(t)$ disappeared above $B_{ext}\approx 1T$ for the device parameters studied here.

The trajectories,  $\vec{r}_i(t)$, were confined to 1 $\mu$m wide channels   (Fig.~1(a)), reflecting the devices used in Refs.~\onlinecite{frolov_prl, frolov_nature}.  The range of spin-orbit parameters and mean free paths explored in this work represent a broader range than would commonly be encountered in a transport experiment.
The channels were assumed to be infinitely long, and each trajectory started from the middle of the wire with a random initial velocity direction.  (It was confirmed that initial conditions had no effect on the calculated spin relaxation times after averaging over an ensemble of trajectories.)

Disorder was taken into account primarily through small-angle scattering, although \disorder~compares the effect of various scattering mechanisms and of soft vs.~hard-wall confinement. With the exception of \disorder, scattering from channel walls was assumed to be specular.  The semiclassical approximation used in this simulation--classical trajectories with coherent spin precession--is valid when both the orbital phase coherence time and the momentum scattering time are shorter than the spin relaxation time, and when electron trajectories can be assumed to be independent of spin direction. The  latter criterion implies that the Fermi energy is much larger than the spin-orbit energy $H_{so}$.  This is a valid approximation in \emph{n}-type GaAs quantum wells but not in \emph{p}-type samples or narrow-gap semiconductors.\cite{Winkler}

\section{Results}

Figure~\ref{traj} shows a short segment of a trajectory defined by small-angle scattering with mean free path of $\lambda=10\mu$m; this level of disorder that is experimentally accessible in high mobility electron gases.  The qualitatively different characteristics of the $x$- and $y$-components of momentum (Figs.~\ref{traj}(b) and ~\ref{traj}(c)) highlight the importance of device geometry in the low-disorder regime studied here.  Electrons  bounce off the channel walls many times before their momentum is randomized, because the mean free path is much larger than the channel width and scattering from the walls is specular.

The  trajectories in such a system are characterized by rapid, nearly-periodic changes in the sign of the momentum transverse to the channel, $k_y$, while the magnitude of the longitudinal momentum, $k_x$, changes only diffusively over a longer timescale.  The square-wave character of $k_y(t)$ can be seen in \traj (c), and in its power spectrum $S_{ky}(f)$ (\traj (d)).  Notice that the $k_y$ frequency spectrum is strongly peaked despite the random angles of electron motion in a typical trajectory (\traj (a)).  The peak frequency in $S_{ky}(f)$ reflects an average over the random distribution of trajectory angles, where the bouncing frequency for angle $\theta$ is $\frac{v_Fcos(\theta)}{2w}$ for channel width $w$.

Relaxation of spins that are aligned initially along $\vec{B}_{ext}$ results from fluctuating fields transverse to $\vec{B}_{ext}$.  In the DP mechanism, those transverse fields are the momentum-dependent effective fields arising from spin-orbit interaction. The first-order component of the effective magnetic field due to $k_y$ is always in the $x$-direction (Eq.~(\ref{eq:bso})), independent of the relative strength of Rashba and Dresselhaus terms in the spin-orbit interaction.  Similarly, the effective field due to $k_x$ is always in the $y$-direction.  Hence spins in an external field along $\hat{x}$ relax due to fluctuations in the motion along $\hat{x}$; spins in a field along $\hat{y}$ relax due to fluctuations in the motion along $\hat{y}$.

\begin{figure} 
 \includegraphics[width= 0.45\textwidth]{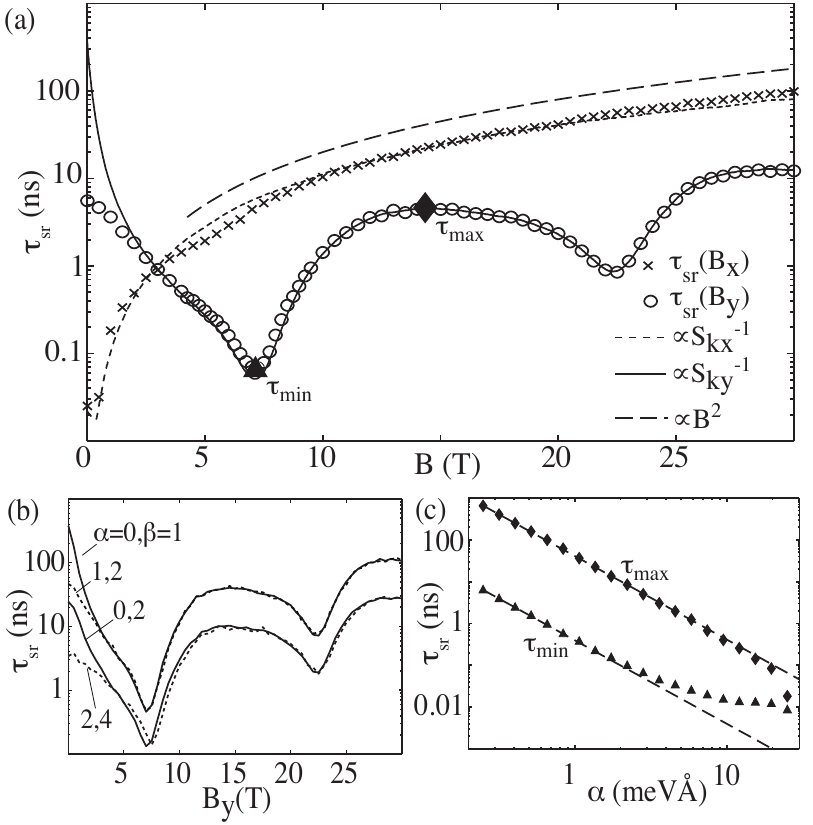}
\caption{\label{soy}  {\bf (a)}  Spin relaxation times for $x$- (crosses) and $y$- (circles) oriented magnetic fields, for trajectories with the same disorder parameters as shown in Fig.~1, $\alpha=3$meV\text{\AA}~and $\beta=\gamma=0$ (see text). Solid and short dashed lines are proportional to ${S_{kx,y}}^{-1}$ as described in the text. Long dashed line shows $B^2$ functional form expected in 2D.  {\bf (b)} Relaxation times for different linear spin-orbit coupling strengths, denoted as ($\alpha,\beta$).  At high field the relaxation time $\tau_{sr}(B_y)$ is determined by $(\alpha - \beta)$, while at low fields $\tau_{sr}(B_y)$ also depends significantly on $(\alpha+\beta)$. 
{\bf (c)} Relaxation time at the first local minimum ($\tau_{min}$, $B_y\approx 7$T) and local maximum ($\tau_{max}$, $B_y\approx 14.5$T), plotted against $\alpha$ ($\beta=\gamma=0$). Throughout the low-spin-orbit range ($\alpha<6$meV\text{\AA}), $\tau_{sr} \propto \alpha^{-2}$  (dashed lines).  }\end{figure}

The qualitatively different power spectral densities of the two momentum components, $S_{kx}(f)$ and $S_{ky}(f)$ (Fig.~1(d)), give rise to qualitatively different relaxation behaviors for spins in $x$- and $y$-oriented fields respectively (Fig.~2(a)).  $\tau_{sr}(B_x)$ increases smoothly with $B_x$ (spins initialized along $\hat{x}$), matching the $\tau_{sr}\propto B_{ext}^2$ behavior expected at high field in 2D disordered systems (Eq.~(\ref{dp2d})) despite the confinement to a micron-wide channel in the simulation.  The sharp periodic dips in $\tau_{sr}(B_y)$ are the BSR features that are the subject of this paper: the short relaxation time at these dips is spin resonance due to the peak frequencies in $S_y(f)$.  This resonance occurs when peaks in $S_y(f)$ occur at the Larmor frequency of the external field, $f_L=g\mu_B B_{ext}/h$.

Figure 2(a) also compares the inverse of the spectral densities of the two momentum components to the spin relaxation times, $\tau_{sr}(B_{ext})$, extracted from the simulations.    Clearly, ${S_{k}}^{-1}(f)$ is directly proportional to $\tau_{sr}(B_{ext})$ when $S_{k}(f)$ is evaluated at the Larmor frequency, $f_L$.  This is reminiscent of the nuclear spin relaxation time, $T_1$, for nuclear magnetic resonance, where it has been shown that ${T_1}^{-1}=(g\mu_B/\hbar)^2 {S_{B_\perp}}(f_L)$, with ${S_{B_\perp}}(f_L)$ representing the spectral density of fluctuations in the transverse magnetic field, $B_{\perp}$, at the Larmor frequency of the static NMR field.\cite{Slichter}    Fluctuations in $B_{so}$ are proportional to $S_{k}(f)$ by Eq.~(2), and it is these fluctuations that lead to relaxation in the present case.  As seen in Fig.~2(a), the approximation $\tau_{sr}(B)\propto {S_{k}}^{-1}(f)$ becomes significantly worse when the mean free path is longer than 10$\mu$m, perhaps because the approximation of exponentially-correlated noise in the NMR result breaks down.

Because $\tau_{sr}(B_y)$ depends on the $x$-component of $\vec{B}_{so}$, it is controlled by $(\alpha-\beta)$ and is nearly independent of $(\alpha+\beta)$ (Eq.~\eqref{eq:bso}).  The accuracy of this approximation can be tested in the simulation by varying $\alpha$ and $\beta$ independently.  As seen in Fig.~2(b), curves with identical $(\alpha-\beta)$ but different  $(\alpha+\beta)$ fall on top of each other for $B_{y}\gtrsim 3$T.    (The stronger dependence on $(\alpha+\beta)$ at low field comes about because the direction of $\vec{B}_{tot}$ fluctuates significantly when $B_{ext}\lesssim B_{so}$.)  The dependence of $\tau_{sr}$ on $\alpha$ is shown in Fig.~2(c), holding $\beta=0$ for all curves:  when examined for particular values of magnetic field, the relation $\tau_{sr}\propto\tau_{so}^2\propto\alpha^2$ expected for 2D (Eq.~(\ref{dp2d})) carries over to the channel data (Fig.~2(c)). 

\begin{figure} 
 \includegraphics[width= 0.45\textwidth]{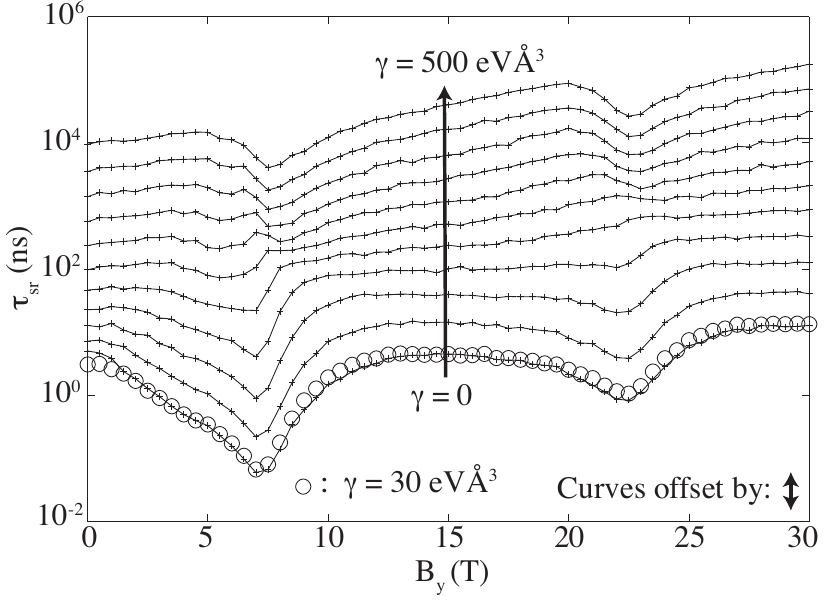}
\caption{\label{gamma} Effect of third-order spin orbit term (quantified by $\gamma$) on BSR features.  Each solid-line curve represents a different $\gamma\in\{0,50,100,...,500\}\text{eV\AA} ^3$, vertically offset as shown.  Circles represent $\gamma=30\text{eV\AA} ^3$  (value expected for GaAs 2DEGs), and fall on top of the $\gamma=0$ curve.  All data is for $\alpha=3\text{meV\AA}$, $\beta=0$, and $\lambda=10\mu $m.} \end{figure}

The discussion thus far has ignored the $O(k^3)$ term in Eqs.~(1) and (2).   Averaged over the Fermi circle, the strength of this term is $\frac{2}{\pi}\gamma k_F^3$.  Using values for  $\gamma$ reported in the literature ($9-34\text{eV\AA} ^3$)\cite{Krich2007} and 2DEG parameters $v_F=1\times 10^5m/s$ and $|\alpha|+|\beta|\approx3\text{meV\AA}$ reported in Ref.~\onlinecite{frolov_nature}, the third-order spin-orbit field, $B_{so}^{(3)}$, is an order of magnitude smaller than the first-order field, $B_{so}^{(1)}$.  For this reason, the simulations presented in most of this paper set $\gamma$ explicitly to zero for ease of calculation.

For significantly larger values of $v_F$ or $\gamma$, on the other hand, $B_{so}^{(3)}$ is of the same order or larger than $B_{so}^{(1)}$.  Because of the more complicated symmetry of $B_{so}^{(3)}$, its effect on BSR  is not monotonic in $\gamma$.  Figure \ref{gamma} explores the role of $B_{so}^{(3)}$ by raising $\gamma$ while holding $v_F=1\times10^5 m/s$, $\alpha=3\text{meV\AA}$, and $\beta=0$.  The BSR dips disappear around  $\gamma\approx300 \text{eV\AA} ^3$, where $B_{so}^{(3)}\approx2\times B_{so}^{(1)}$, but then revive for larger values of $\gamma$.    
  
 \begin{figure} 
 \includegraphics[width= 0.45\textwidth]{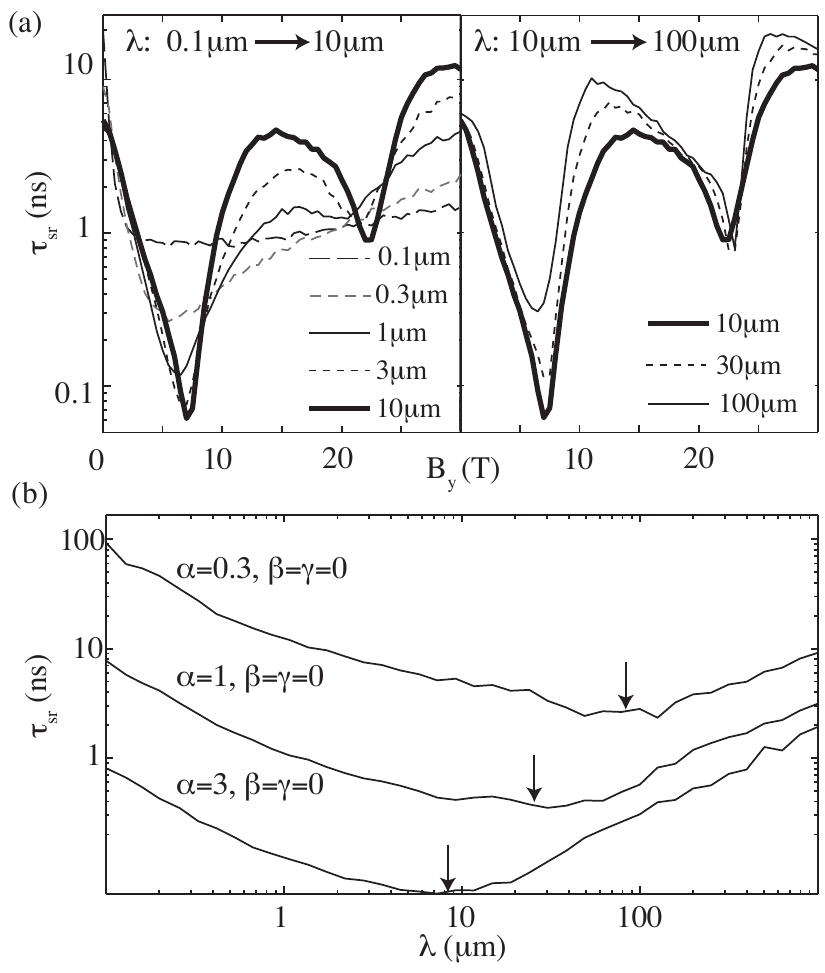}

\caption{\label{mfp} {\bf (a)} Relaxation times for different mean free paths in a 1$\mu$m-wide channel.  Left- and right-hand panels show mean free paths less than and greater than $10\mu$m respectively, with the $\lambda=10\mu$m trace appearing in both panels.  Signature of ballistic spin resonance is clearly visible even when $\lambda$ is as short as the channel width $1\mu$m. Features continue to sharpen (including faster spin relaxation at resonance) up to $\lambda=10\mu$m.  For longer mean free paths, relaxation time at resonance rises again. {\bf (b)} Effect of mean free path on relaxation time at the first local minimum ($\tau_{min}$, $B_y\approx 7$T) in a 1$\mu$m-wide channel.  Arrows denote the crossover value $\lambda_{min}=2\pi v_F \tau_{so}$ discussed in text.} 
 \end{figure}

Significant changes in spin relaxation were observed when the overall magnitude of disorder (set by $\lambda$) was changed (\mfp).  When $\lambda$ was much smaller than the channel width, resonant dips were absent.  In that case, the bouncing frequency ceases to be a relevant parameter, as electrons seldom make it across the channel without scattering, and the 2D limit of Eq.~(\ref{dp2d}) is approached.  The dips become deeper as $\lambda$ is increased, but reach a minimum value around $10\mu$m before rising again for even longer mean free paths.

In order to understand this non-monotonic dependence, we study the $\lambda$-dependence of $\tau_{sr}$ at the first resonant dip, around $B_y=7T$ (Fig.~4(b)).   Starting from very short mean free paths, $\tau_{sr}$ reaches a minimum at $\lambda_{min}= 2\pi v_F \tau_{so}$, then rises again for very long mean free path.   The length-scale, $\lambda_{min}$, corresponds to the distance an electron would have to travel in order for the spin to rotate by $2\pi$ due to the spin-orbit effective field.

This behavior can be explained at a qualitative level by considering spin relaxation in a reference frame that rotates at the Larmor frequency.  Working in this frame effectively removes precession due the external field, {\em and} it removes flips in $\vec{B}_{so}$ that occur at frequency $f_L$ due to bouncing between the channel walls.  In other words, spin relaxation in the ballistic channel at the BSR condition is approximately mapped onto spin relaxation in a disordered 2D system at zero external magnetic field.  In 2D at zero external field, one expects $\tau_{sr}\sim\tau_{so}^2/\tau_p\equiv \tau_{so}^2v_F/\lambda$ (Eq.~(3)) to decrease with increasing $\lambda$ in the motional narrowing regime, i.e.~for fast momentum relaxation, $\tau_p < \tau_{so}$.  In the ballistic limit, $\tau_p > \tau_{so}$, on the other hand, one expects $\tau_{sr}$ to increase with $\lambda$ as $\tau_{sr}\sim\tau_p=\lambda/v_F$ because spins precess coherently between scattering.\cite{Kiselev:2000, Koop}  It is the crossover from motional narrowing to ballistic regimes that gives rise to the non-monotonic behavior of $\tau_{sr}$ in Fig.~4(b).

\begin{figure} 
\vspace{0.1in}
 \includegraphics[width= 0.45\textwidth]{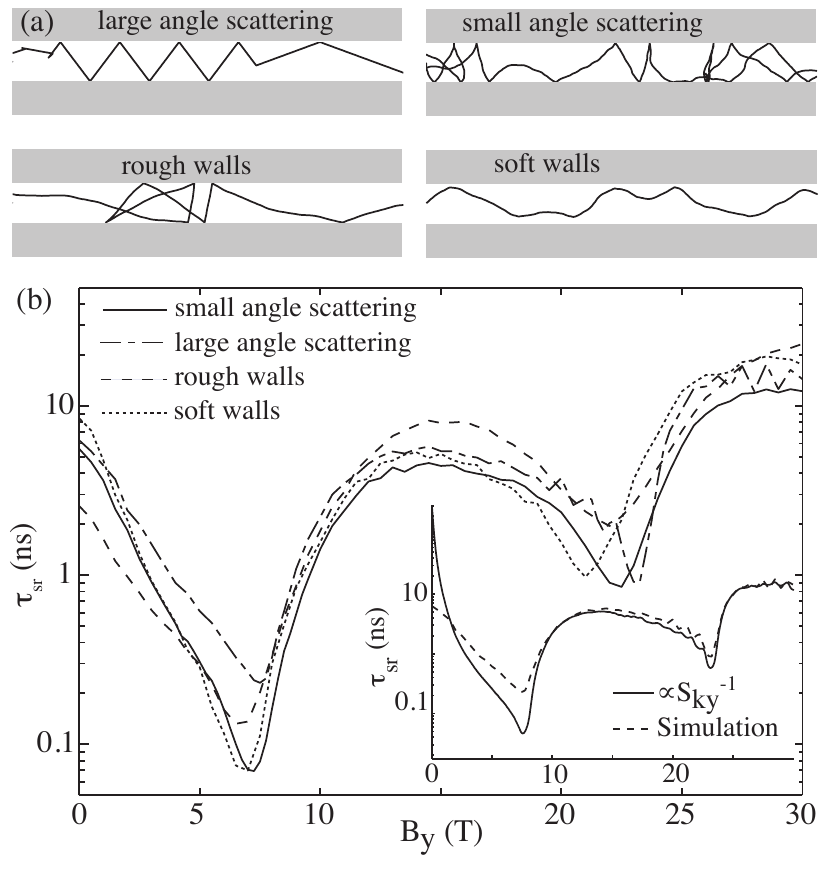}

\caption{\label{disorder} {\bf (a)} Cartoon trajectories corresponding to the types of disorder and scattering described in the text. {\bf (b)} Spin relaxation time for $B_{ext}\parallel\hat{y}$, calculated for four disorder models, each giving effective $\lambda$=10$\mu$m.  Solid: small angle scattering.   Dash-dot: large angle scattering. Short dash: rough walls (non-specular scattering), plus small angle scattering giving $\lambda$=100$\mu$m for trajectories that would otherwise not encounter a wall.  Dotted: small angle scattering and soft walls, modelled by a carrier density that decreases smoothly to zero within $\lambda_{dep}=150$nm of channel walls.  Inset: match between the power spectral density and spin relaxation time is much worse in the case of large angle scattering, compared to small angle scattering (see, e.g., Fig.~2(a)) and other disorder models (data not shown).  ($\alpha=3$meV\text{\AA}, $\beta=\gamma=0$)} \end{figure}

Finally, we show that the particular type of disorder used to generate trajectories, and the type of scattering off channel walls, has only a small effect on the simulated spin relaxation curves.  Figure \ref{disorder} shows spin relaxation for three different types of disorder: 

\begin{enumerate}

\item {\em small-angle scattering}. The direction of motion changed from timestep to timestep by a small angle that was Gaussian-distributed around zero, with standard deviation calculated to give the desired mean free path.  This is believed to be the dominant scattering mechanism in high-mobility GaAs 2DEGs.\cite{Coleridge:1991, juranphys} 

\item {\em large-angle scattering} was implemented as a probability for complete randomization of momentum angle at each timestep.  The probability was calculated to give the desired mean free path.

\item {\em rough potential walls} Upon reflection off channel walls, the angle of reflection was randomly distributed around the angle of incidence with a spread of $\phi_{spec}$.  $\phi_{spec}=0$ corresponds to specular scattering from channel walls.  This effect is believed to be weak in electrostatically-defined GaAs 2DEG nanostructures, as shown by clear transverse focusing signals even up to high order, which require many specular bounces.\cite{VanHouten:1989}

\end{enumerate}

Each curve shown in Fig.~5(b)  corresponds to disorder from only one of the three mechanisms.  The mean free path is $\lambda=10\mu$m in each case, confirmed by monitoring the autocorrelation of $k_x(t)$.  As seen in the figure, the simulated spin relaxation time depends only slightly on the precise model of disorder, despite the importance of ballistic transport to the resonant dips in $\tau_{sr}(B_y)$.  Figure 5(b) also compares BSR for the case of simple reflections from hard-wall channel boundaries to the more realistic case of soft walls, with a 150nm depletion length as might be expected in nanostructures defined by electrostatic surface gates.  Small angle scattering is implemented to give $\lambda=10\mu$m in both hard-and soft-wall simulations.   The difference between the hard- and soft-wall data is nearly indistinguishable, except for a small shift in the field at which the resonance dips occur.

{\small {\bf Acknowledgements:}  The authors thank J.C.~Egues, M. Lundeberg, and G.~Usaj for valuable discussions. Work at UBC supported by NSERC, CFI, and CIFAR.}

 \end{document}